# Algorithms for Generating Large-scale Clustered Random Graphs

Abstract

Real social networks are often compared to random graphs in order to assess whether their typological structure could be the result of random processes. However, an Erdős-Rényi random graph in large scale is often lack of local structure beyond the dyadic level and as a result we need to generate the clustered random graph instead of the simple random graph to compare the local structure at the triadic level. In this paper a generalized version of Gleeson's algorithm is advanced to generate a clustered random graph in large-scale which persists the number of nodes $|V|$, the number of edges $|E|$, and the global clustering coefficient $C_\Delta$ as in the real network. And it also has advantages in randomness evaluation and computation time when comparing with the existing algorithms.

1. Introduction

Random graph models are usually being used to compare with real networks. The random graph preserves the number of nodes $|V|$ and the number of edges $|E|$ of the real network, and it does work for a small network with hundreds of or thousands of nodes and thousands of or tens of thousands of edges. However, as the network size grows larger and larger, an Erdős-Rényi random graph fails to reproduce the local structure beyond dyadic level which is correlated with non-zero clustering coefficient, "small world" phenomenon, and other important network characteristics.

There are at least four algorithms advanced to generate a random graph with clustering since 2009. However, none of these algorithms has been tested for large-scale networks. In this paper I will go over these algorithms, examine their feasibility, advantages, and disadvantages, and make some revisions if necessary for constructing a clustered random graph in large scale.

2. Network/graph notations and some basic properties

A graph – usually denoted as $G(V, E)$ or $G = (V, E)$ – consists of a set of vertices, nodes, egos, agents, or objects $V$, and a set of edges, links, ties, arcs, or adjacency relations $E$. Edges are often represented by $e_{ij}$ or $e = (i, j)$ where $i$ and $j$ denote the ends of $e$. If each edge $e_{ij}$ in a graph $G(V, E)$ is defined by the unordered pairs of nodes $i$ and $j$, $G$ is regarded as an undirected graph, which means the relationship between its end points $i$ and $j$ is symmetric $e_{ij} == e_{ji}$. If each edge $e_{ij}$ in a graph $G(V, E)$ is defined by the ordered pairs of nodes $i$ and $j$, $G$ is regarded as a directed graph, which means the relationship between its end points $i$ and $j$ is asymmetric $e_{ij} \neq e_{ji}$.

In the undirected graph, the nodal degree $d_i$ is the number of edge(s) incident upon a node $i$. In a directed graph the number of edges sending out from a node is called the indegree $d_{i\_in}$ and the number of edges received by the nodes is its outdegree $d_{i\_out}$ for a node $i$.

The mean nodal degree or average degree of a graph is a statistic that reports the average degree of the nodes in the graph, which can be calculated as

$$\bar{d} = \frac{|E|}{|V|}$$

Network density is the proportion of possible lines that are actually present in the graph, which can be calculated as

$$\Delta = \frac{|E|}{|V|(|V|-1)/2} = \frac{2|E|}{|V|(|V|-1)}.$$

3. Transitivity and clustering coefficient

It is commonly said "a friend of a friend is a friend" (Wasserman and Faust, 1994, p.150). At the beginning of 20th century, Simmel (1908/1950) argues that triad is the fundamental social unit and a strong social tie could not exist without being part of a clique. In other words, a person is more likely to share contacts with her or his close contacts, and close contacts have an increased likelihood of knowing each other. Triadic closure is also studied by Heider (1946), Cartwright and Harary (1956), Davis (1967, 1979), Granovetter (1973), Krackhardt (1998), Krackhardt and Handcock (2006), and Opsahl and Panzarasa (2009).

In an undirected graph, a relationship is transitive if node $i$ is adjacent to node $j$ and node $j$ is adjacent to node $k$, node $i$ is also adjacent to node $k$. The property in a graph $G$ can be measured by transitivity or clustering coefficient. The clustering coefficient is a measure of degree to which nodes in a graph $G$ tend to cluster together.

The local/individual clustering coefficient of an undirected graph gives an indication of the embeddedness or cohesiveness of a single node. Based on Watts and Strogatz's work (1998), for a node $i$, its local clustering coefficient $C(i)$ is the ratio of number of triangles[1] it involves in, $\Delta(i)$, to the number of possible triangles it could involve in, $\tau(i)$. The number of triangle(s) a node $i$ involves in is given by

$$\Delta(i) = \left| e_{ik} \in E \mid e_{ij} \in E \ \& \ e_{jk} \in E \right|$$

The number of possible triangles a node $i$ could involve in is depending on her or his nodal degree $d_i$ and can be represented as

$$\tau(i) = \frac{d_i(d_i - 1)}{2}$$

And thus the local clustering coefficient of a node $i$ whose degree is greater than two can be represented as

$$C(i) = \frac{\Delta(i)}{\tau(i)} = \frac{2\Delta(i)}{d_i(d_i - 1)}$$

The average clustering coefficient $C(G)$ of a graph $G$ is defined as the average of the local clustering coefficient of its nodes $V$. More precisely

$$C(G) = \frac{\sum_i C(i)}{|V|}$$

The global/overall clustering coefficient $C_\Delta$ or transitivity measure $T(G)$ reflects the overall clustering property in the graph. It is defined as the ratio of three times the number of triangles in the graph, $\Delta(G)$, to the number of 2-paths[2] in the graph, $\tau(G)$. That is

$$C_\Delta = T(G) = \frac{3\Delta(G)}{\tau(G)}$$

---

[1] Triangle refers to a network structure of three nodes which connect with one another.
[2] 2-path refers to a network structure that an ego has two alters. If these two alters are connected, it is a triangle; and if not, it is a structural hole.

The difference between the average clustering coefficient $C(G)$ and the global/overall clustering coefficient $C_\Delta$ or transitivity measure $T(G)$ is that the former is the mean of the sum of mean – the local clustering coefficient $C(i)$ is a mean measure and it is summed up for all the nodes $i$ in the graph $G$ and divided by the number of nodes $|V|$ in the graph – while the latter is a mean measure of sum – it is calculated by the sum of the closed 2-paths over the total number of 2-paths. And according to mathematical logic we know that the sum of mean is not equal to the mean of sum

$$\frac{a_1}{b_1} + \frac{a_2}{b_2} + \cdots + \frac{a_n}{b_n} \neq \frac{a_1 + a_2 + \cdots a_n}{b_1 + b_2 + \cdots b_n},$$

and thus the average clustering coefficient $C(G)$ and the global/overall clustering coefficient $C_\Delta$ or transitivity measure $T(G)$ are not identical to each other. Usually both clustering coefficients are measured at the same time in network study.

4. Data and network properties

The data I use in this paper come from a mobile phone network of over 10 million subscribers of one unnamed mobile phone company[3], and the raw data provide details of time, origins, call types, destinations and durations. We focus on the voice-call communication behaviors during 4 weeks – from August 3, 2008 (Sunday) to August 30, 2008 (Saturday) and convert it to an undirected graph.

---

[3] The network data have been used in numerous publications including Bagrow et al. (2011); Ercsey-Ravasz et al. (2011); Ghoshal and Barabási (2011); Hidalgo and Rodriguez-Sickert (2008); Lichtenwalter et al. (2010); Liu et al. (2011); Onnela et al. (2011); Raeder et al. (2011); and Wang et al. (2011).

This piece of network data consist 6,719,330 active nodes[4] and 15,913,611 edges[5]. And thus the average nodal degree is $\bar{d} = \frac{|E|}{|V|} = \frac{15,913,611}{6,719,330} = 2.37$ and the network density is

$$\Delta = \frac{2|E|}{|V|(|V|-1)} = \frac{2 \times 15,913,611}{6,719,330 \times (6,719,330-1)} = 7.05e-7.$$

At the triadic level, there are 126,175,382 2-paths among which 109,383,149 are structural holes[6] and 5,597,411 are triangles. The average clustering coefficient $C(G) = \frac{\sum_i C(i)}{|V|}$ is 0.24. The global/overall clustering coefficient $C_\Delta$ or transitivity measure $T(G)$ is

$$\frac{3\Delta(G)}{\tau(G)} = \frac{3 \times 5,597,411}{126,175,382} = 0.13.$$

5. Random graph models

The random graph model is usually applied to compare with the property of real network. It is generated by adding edges between a set of *n* nodes at random. The first random graph model is proposed by Erdős and Rényi (1959), denoted as $G(n, p)$, which has *n* nodes (identical to |E|) and each possible edge comes into being independently with the probability $p \in (0, 1)$.

Several characteristics of Erdős-Rényi model can be extracted. For example, when the network grows larger and larger, the average nodal degree in the network is constant as $\bar{d} = \dfrac{p(n-1)}{2}$ and the graph has a Poisson degree distribution: the probability that a node in the network has the degree $d_i$ is $p_{d_i} \simeq \dfrac{\bar{d}^{d_i} e^{-\bar{d}}}{d_i!}$; the expected number of edges is $\dfrac{n(n-1)}{2} p$; and since the probability that two nodes to form an edge is $p$ which is independent from whether they share a common neighbor or not, as the network size increases, the average clustering coefficient of the network $C(G)$ approaches 0.

---

[4] The total number of customers is about 10 million and about 6.7 million of them have at least one communication behavior during the four weeks.
[5] In the undirected graph the relationship between any two nodes *i* and *j* is symmetric $e_{ij} == e_{ji}$, and as a result I can use either double counting – both the $e_{ij}$ and $e_{ji}$ are included in the edge list – or single counting – only one of $e_{ij}$ and $e_{ji}$ is included in the edge list – and the number of edges |E| in the former strategy is twice as that of the latter one. In this study I adopt single counting and all the formulas are adjusted for this situation.
[6] Structural hole is first advanced by Burt (1995) and refers to a structure that an ego has two alters who does not connect with each other.

Since we also have $\bar{d} = \dfrac{|E|}{|V|}$ we can infer that the probability $p$ of the formation a tie between any two nodes in a large-scale network is equal to the network density

$$p = \dfrac{2\bar{d}}{n-1} = \dfrac{2\dfrac{|E|}{|V|}}{|V|-1} = \dfrac{2|E|}{|V|(|V|-1)} = \Delta.$$

6. Random graph models with clustering

Random graph models are supposed to reproduce several properties of real networks, including the number of nodes |V| and the number of edges |E|, and if it is a small network with hundreds of or thousands of nodes and thousands of or tens of thousands of edges, we can compare other network properties at all levels between the random graph and the real network to see how non-random the real network is.

However, by preserving the number of nodes |V| and the number of edges |E| of the real network, the random graph models only successfully reproduce the characteristic of skewed degree distribution – most nodes have low degrees but a small number, known as "hubs", have high degrees. When the network size grows as large as in Facebook, Twitter, or in a mobile phone network, the situation is quite different. In earlier section we mention that as the network

size grows larger and larger, the average clustering coefficient in a random graph approaches zero, which means in a large-scale simple random graph the existence of triangles or higher-order local structures is really rare by chance. And thus it is meaningless to compare the local structure beyond dyadic level since there is no such a thing in a random network in large scale with zero clustering coefficient. What's more, there is a "small world" (Travers & Milgram 1969) or "six degree of separation" (Watts & Strogatz 1998) phenomenon in the real world, in other words, the nodes in the real network have limited geodesic distance. However, in the large-scale simple random graph the edges are scattered by chance and the geodesic distance will be infinity.

That is why we need to generate the random graph with clustering. Not only do we fix the number of nodes $|V|$ at the nodal level and the number of edges $|E|$ at the dyadic level, but we try to fix the average clustering coefficient $C(G)$ or global clustering coefficient $C_\Delta$ at the triadic level. In this way we can reproduce the characteristics of non-zero clustering coefficient and limited geodesic distance as in the real network. And the characteristics of non-zero clustering coefficient and limited geodesic distance are also correlated with other important network properties such as the component sizes, the existence and size of a giant component, and the percolation properties. And it will also enable us to study the network robustness, percolation, and cascading failure, the diffusion process, and the effect of network topology on the dynamical systems. There are at least four algorithms to generate random graph with clustering advanced in recent years and we will go over each one of them in the following paragraphs.

6.1 Algorithm of Guo and Kraines

One of the models used to generate random graph with clustering is presented by Guo and Kraines in 2009. The algorithm is composed of three steps. First, a random graph $G$ with a set of nodes $V$ and a set of edges $E$ is constructed following the given power-law degree distribution[7]. Second, five nodes in the random graph are randomly selected and two edges among them are partly rewired to add one triangle. Third, the process will be repeated until either the average clustering coefficient of the rewired graph is greater than or equal to the target average clustering coefficient $C(G)$, or it reaches a certain predefined number of trials.

As Figure 1 shows, in the second step the five nodes randomly picked from the random graph should meet the following conditions:

1) $x$ and $w$ are alters of $v$;

2) $y$ and $z$ are <u>not</u> alters of $v$;

3) $e_{wy}$ and $e_{xz}$ do exist.

4) $e_{wx}$ and $e_{yz}$ do **NOT** exist.

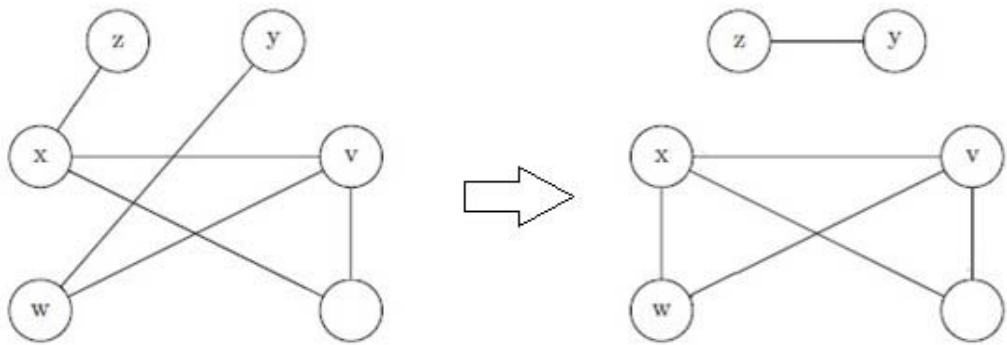

Figure 1. Algorithm of Guo and Kraines (Guo and Kraines, 2009)

---

[7] Power-law degree distribution is one of the skewed degree distribution modes. Barabási and Albert (1999) argue that a common property of many large networks is that the vertex connectivities follow a scale-free power-law distribution and this feature was found to be a consequence of two generic mechanisms: (1) networks expand continuously by the addition of new vertices, and (2) new vertices attach preferentially to sites that are already well connected.

As Figure 1 shows the two edges are rewired from $e_{wy}$ and $e_{xz}$ to $e_{wx}$ and $e_{yz}$ and one triangle among $v$, $w$, and $x$ is added.

All the rewiring processes start from a random graph $G(V, E)$ with zero clustering coefficient. The average clustering coefficient $C(G)$ will be updated each time when the two pairs of randomly-selected edges are rewired until it reach 0.24 – the average clustering coefficient $C(G)$ in the real network. And to update the average clustering coefficient $C(G)$ we need to refresh both the numerator $\Delta(i)$ – the number of triangles each node $i$ involves in and the denominator $\tau(i)$ – the number of possible triangles each node $i$ could involve in. We have mentioned before that the average clustering coefficient $C(G)$ is the mean of sum of mean. It works fine for a small network with hundreds or thousands of nodes (i.e., Guo and Kraines generated a test network of 1000 nodes and 3000 edges) but is not feasible for a large-scale network with millions of nodes and tens of millions of edges since it will take unacceptable long time to update the triangle list and 2-path list millions of times.

An alternative plan is to substitute the average clustering coefficient $C(G)$ with the global clustering coefficient $C_\Delta$ or transitivity measure $T(G)$ – the mean of sum. In this case it is 0.13 in the real network.

We should notice that in step 1 the initiated random graph $G$ has fixed number of nodes $|V|$ and edges $|E|$ as in the real network but not the fixed nodal degree $d_i$ for each node $i$, and as a result the number of 2-paths in the random graph, $\tau'(G) = \sum_i d'_i(d'_i - 1)/2$, is not the same as in the real network. But in step 2 the rewiring process keeps the nodal degree $d'_i$ for each node $i$ as in the initiated random graph $G$, as a result the same number of 2-paths in the graph $\tau'(G)$ be persisted after that.

There are two advantages for adopting the global clustering coefficient $C_\Delta$ or transitivity measure $T(G)$. First, since the number of 2-paths $\tau'(G)$ is consistent as in the initiated random graph $G$ and we know the global clustering coefficient $C_\Delta$ or transitivity measure $T(G)$ in the real network is 0.13, the expected global clustering coefficient $C_\Delta$ or transitivity measure $T(G)$ is simplified as the expected number of triangles $\Delta(G) = \tau'(G) \times 0.13/3$ and we know each rewiring process adds one triangle to the graph and we just need to do $\Delta(G)$ times of rewiring. Second, the rewiring processes can be done in batch and that will save a lot of computation time.

In the first step we generate a random graph $G$ with 6,719,330 nodes and 15,913,611 edges. The average network degree of the initiated random graph $G$ is

$$\bar{d} = \frac{|E|}{|V|} = \frac{15,913,611}{6,719,330} = 2.37$$

and the network density is $\Delta = \dfrac{2|E|}{|V|(|V|-1)} = \dfrac{2 \times 15,913,611}{6,719,330 \times (6,719,330-1)} = 7.05e-7$.

But the total number of actual 2-paths in the initiated random graph $G$ is

$$\tau'(G) = \sum_i d_i'(d_i' - 1)/2 = 72,017,211.$$

And thus the expected number of triangles in the initiated random graph $G$ is

$\Delta(G) = \tau'(G) \times C_\Delta / 3 = \tau'(G) \times (5,597,411 \times 3/126,175,382) / 3 = 3,194,838$.

There are 54 triangles in the initiated random graph $G$, which is very close to the number of 2-paths times the tie formation probability $p$ in the random graph, the value of which is equal to that of the network density $\Delta$

$$72,017,211 \times \frac{2 \times 15,913,611}{6,719,330(6,719,330-1)} = 51.$$

After rewiring 3,194,784 = 3,194,838 (the expected number of triangles ) - 54 (the number of triangles in the initiated random graph $G$) times, we get a clustered random graph with 3,195,086 triangles. And the global clustering coefficient $C_\Delta$ or transitivity measure $T(G)$ is 3 × 3,195,086/72,017,211 = 0.13, which is the same as in the real network.

The average clustering coefficient of the clustered random graph is 0.22, which is a little bit smaller than that in the real network 0.24.

The rewiring processes are performed in 5 batches and it takes about 490 hours to get the expected clustered random graph on a server with Linux 2.6.18-274.12.1.e15 operating system, two Intel Xeon X5450 3.00GHz 4-core CPUs, 64GB DDR2 667MHz PC2-5300 RAM, and twelve Western Digital WD1001FALS-0 hard drives (7200 RPM, 1TB, 32MB cache) in a RAID 60 array.

6.2 Algorithm of Bansal et al.

Bansal et al. (2009) advanced a second random graph generator with clustering. The algorithm consists of 3 steps. First, the real network is rewired to be a completely random graph with same number of nodes $|V|$, same number of edges $|E|$, and same nodal degree $d_i$ for each node $i$.

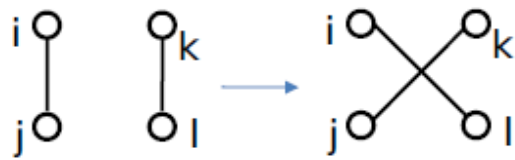

Figure 2. Edge rewiring in step 1 of algorithm of Bansal et al. (Bansal et al. 2009)

As showed in Figure 2, four nodes are randomly selected from the real network that should meet the following conditions:

1) $e_{ij}$ and $e_{kl}$ do exist.

2) $e_{il}$, $e_{ik}$, $e_{jk}$, and $e_{jl}$ do **NOT** exist;

The edges $e_{ij}$ and $e_{kl}$ are rewired to the edge $e_{il}$ and $e_{ik}$ and the process will be repeated until the global clustering coefficient $C_\Delta$ or transitivity measure $T(G)$ is equal to the density, which means the closing off of the 2-paths in the graph is totally a random process and its probability is just equal to the independent probability of an edge formation.

Second, as showed in Figure 3, we can pick a chain of five nodes $k, j, i, l$, and $m$ and add one triangle at a time by rewiring $e_{jk}$ and $e_{lm}$ to $e_{jl}$ and $e_{km}$. Or we can pick a ring of six nodes $i, j, k, n, m$, and $l$ and add two triangles at one time by rewiring $e_{jk}$ and $e_{lm}$ to $e_{jl}$ and $e_{km}$.

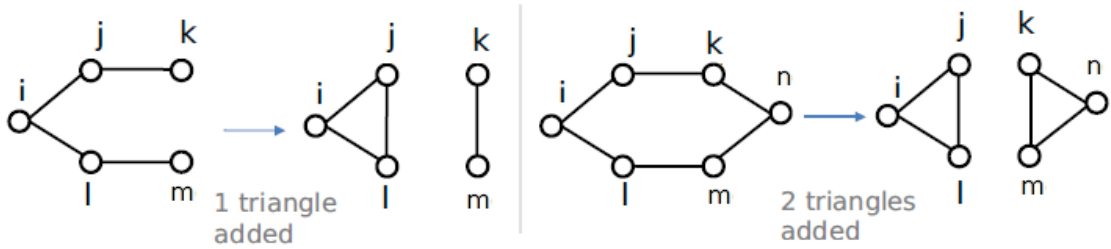

Figure 3. Adding triangles in step 2 of algorithm of Bansal et al. (Bansal et al. 2009)

Third, the process will be repeated until we get both the same transitivity measure $T(G)$ and average clustering coefficient $C(G)$ as in the real network.

The algorithm of Bansal et al. can be seen as an upgraded version of that of Guo and Kraines. First, in step 1 they keep the nodal degree $d_i$ for each node $i$ by rewiring the real network and thus the number of 2-paths $\tau(G) = \sum_i d_i(d_i - 1)/2$ in the graph is kept (in this case it is 126,175,382), and in step 2 when the transitivity measure $T(G)$ is the same as in the real network (in this case it is 0.13), the expected number of triangles $\Delta(G)$ is also the same (in this case it is 5,597,411). Second, Bansal et al. used both transitivity measure $T(G)$ and average clustering coefficient $C(G)$ as threshold of rewiring in step 2 and step 3 while Guo and Kraines only adopt the average clustering coefficient $C(G)$. And third, they invent a method to add two triangles at a time to save the computation time.

Bansal et al. evaluate their algorithm by comparing several clustered random networks with their correspondent real ones among which the largest network size is 4,713. However, for a large-scale network of 6.7 million nodes and 15.9 million edges, if we want to generate a clustered random network in a reasonable time we should discard the measure of average clustering coefficient $C(G)$ for the same reason when we discuss the algorithm of Guo and Kraines and just try to fit the transitivity measure $T(G)$, which is identical to fit the expected number of triangles $\Delta(G)$. In this way we can rewire the edges in batch and save a lot of time.

We should notice that after the real network is rewired to be totally random in step 1, the expect number of triangles in the rewired network is equal to the number of 2-paths times the tie formation probability $p$ in the random graph, the value of which is equal to that of the network density $\Delta$

$$126,175,382 \times \frac{2 \times 15,913,611}{6,719,330(6,719,330-1)} = 89$$

There are 92 triangles after the real network is rewired 20 multiply $|E|$ times in step 1, which is very close to expected number 89.

After rewiring 5,597,319 = 5,597,411(expected number of triangles) - 92 (the number of triangles in the rewired random graph $G'$) times in step 2, we get a clustered random graph $G''$ with 5,597,632 triangles. And the global clustering coefficient $C_\Delta$ or transitivity measure $T(G)$ is $3 \times 5,597,632/ 126,175,382= 0.13$, which is the same as in the real network.

The average clustering coefficient of the clustered random graph is 0.41, which is much larger than that in the real network 0.24.

The rewiring processes are performed in 27 batches and it takes about 3,150 hours to get the expected clustered random graph on the same server we work for the algorithm of Guo and Kraines.

6.3 Newman's algorithm

Newman (2009) designed a third algorithm to generate a random graph with non-zero clustering coefficient. Newman's algorithm can be understood as getting a clustered random graph $G(V, E)$ from configuration model $G(V, S, T)$ which takes a triangle vector (also called corners which represents the number of triangle(s) a node $i$ is involved in) $t_i$ and a single edge vector (also called stubs which represents the number of edge(s) a node $i$ involves in structural holes) $s_i$ for each node $i$ in the real network as an input. Since the nodal degree of a node $i$ can be represented as $d_i = 2t_i + s_i$, the generating function for the joint distribution of triangles and single edges is given by

$$g(x, y) = \sum_{s,t} p_{st} x^s y^t$$

where $p_{st}$ is the probability that a node involves in $s$ single edges and $t$ triangles.

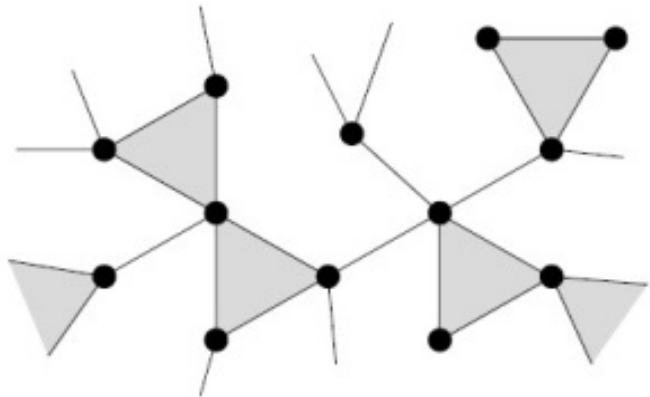

Figure 4. Newman's algorithm (Newman 2009)

Based on Newman's algorithm, as showed in Figure 4 each node $i$ can be viewed as connecting to $s_i$ stubs and $t_i$ corners. Then a triangle is added by joining three corners of nodes $i$, $j$, and $k$ at random and this process is repeated until all the corners are parts of some unique triangles. And a single edge is added by joining two stubs of nodes $i$ and $j$ at random and this process is repeated until all the stubs are parts of some unique single edges.

Residual single edge vector $rs_i$ and residual triangle vector $rt_i$ are used to track the unconnected number of single edges and triangles for each node $i$. They start with the values of $s_i$ and $t_i$ and will minus one when one single edge or triangle is added respectively. At the end of the process, the residual single edge vector $rs_i$ and residual triangle vector $rt_i$ should both be equal to 0 for each node $i$.

Newman suggests that by persisting the number of nodes $|V|$, the number of edges $|E|$, the number of triangles $T$, and the number of single edges $S$, we will get a random graph with the same global clustering coefficient $C_\Delta$ or transitivity measure $T(G)$ as in the real network.

To complete step 1 of randomly connecting three corners to fit the expect number of triangles is much faster on the same server than the first two algorithms. The computation time is only about 3.5 hours. However, the problem is that in the real network the 5,597,411 triangles only use up 8,474,226 edges (about 53.25% of all edges), but by adopting Newman's algorithm the 5,597,411 triangles use 15,171,585 edges (about 95.34% of all edges) in the first step and there are only 742,026 edges left for single edges, which means there will not be enough structural holes generated and thus Newman's algorithm fails to fit the global clustering coefficient $C_\Delta$ or transitivity measure $T(G)$ in the real network.

The problem of Newman's algorithm lies in the fact that it over-uses the edges to produce the same number of triangles as in the real network. In the first two algorithms, the numbers of edges $|E|$ incident to a node $i$ and 2-paths $\tau(G)$ are fixed in the rewiring processes and each time when we rewire 2 edges to add a triangle we know it is one step to get closer to the expected global clustering coefficient $C_\Delta$. However in Newman's algorithm we only focus on adding triangles in step 1 but lose control of the number of edges |E| incident to a node *i* and the number of 2-paths $\tau(G)$, and as a result the triangles are spreading all over the network space. After the 742,026 edges are added to link the 5,597,411 triangles, we only get 91,406,099 2-paths, the global clustering coefficient $C_\Delta$ goes up to 0.18 and the average clustering coefficient $C(G)$ goes up to 0.44.

6.4 Gleeson's algorithm

Gleeson (2009) argues that in Newman's algorithm, if a node *i* has nodal degree $d_i$, then it could be a member of up to $d_i/2$ disjointed triangles and the upper limit of the local clustering coefficient is $C(i) = \frac{d_i/2}{\binom{d_i}{2}} = \frac{1}{d_i - 1}$, and as a result the constraint inhibits its fit to most real networks in the world.

Gleeson (2009) alternatively proposes the γ-theory networks to generate random graph with non-zero clustering coefficient. Instead of joining triangles and single edges incident to each node *i*, he generalizes Newman's model by starting from other higher-order motif – a *k*-clique, which is a complete graph among *k* nodes each of which is connected to every other node in the graph. The joint degree distribution $\gamma_{d_i,k}$ specifies the probability that a randomly selected nodes *i* has degree $d_i$ and is part of a *k*-clique. If a *k*-clique has *k* nodes, each node *i* in the *k*-clique has the same nodal degree $d_i = k - 1$ and the same local clustering coefficients $C(i) = (k - 2)/k = (d_i - 1)/(d_i + 1)$.

Gleeson (2009) uses external links (which is similar to Newman's single edge and represents the edges not involved in any cliques) to join all the *k*-cliques together. For example, if the mean degree of a real network is between 3 and 4, a clustered random graph can be generated by joining some 3-cliques (triangles), some 4-cliques, and with the remainder as individuals (i.e., 1-cliques) as showed in Figure 5.

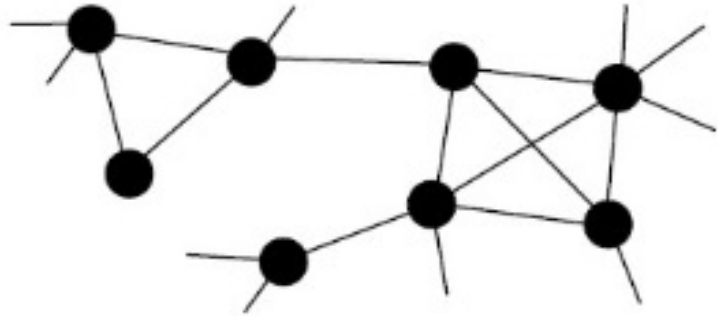

Figure 5. Gleeson's algorithm (Gleeson 2009)

Gleeson calls any nodes in a clique as super-nodes. In Gleeson's algorithm, all the clustering exists among the super-nodes of *k*-cliques and structural holes are formed between each pair of *k*-cliques. If we consider each *k*-clique as a super-super-node, we can reduce the whole network as a random graph with zero clustering coefficient.

Gleeson's algorithm seems to be really promising. And when we apply Newman's algorithm to generate clustered random graph, the triangles over-use the edges and as a result both the global clustering coefficient $C_\Delta$ and the average clustering coefficient $C(G)$ go up. If Gleeson's idea works, one edge will help to generate more triangles in a *k*-clique ($k > 3$) than in a 3-clique – a triangle.

Let's suppose that the expected clustered random graph is composed of three kinds of *k*-cliques – 3-clique, 4-clique, and 5-clique. As showed in Table 1, the 3-clique has three nodes, three edges, and one triangle. The 4-clique has four nodes, six edges, and four triangles. And the 5-clique has five nodes, ten edges, and ten triangles. The expected number of triangles is 5,597,411, the number of edges evolving in the triangles is 8,474,226, and the number of nodes whose nodal degrees are greater than 2 is 5,358,175. And we get

Table 1. Basic statistics of *k*-cliques and some other motifs

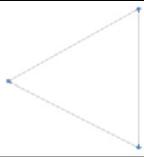 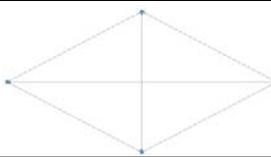 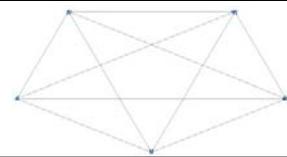

| | | | |
|---|---|---|---|
| # of nodes | 3 | 4 | 5 |
| # of edges | 3 | 6 | 10 |
| # of triangles | 1 | 4 | 10 |
| # of structural holes | 0 | 0 | 0 |

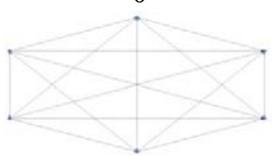 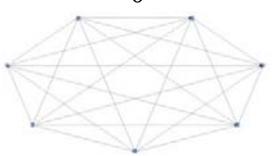 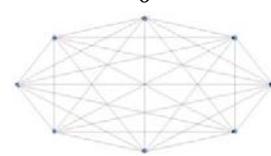

| | | | |
|---|---|---|---|
| # of nodes | 6 | 7 | 8 |
| # of edges | 15 | 21 | 28 |
| # of triangles | 20 | 35 | 56 |
| # of structural holes | 0 | 0 | 0 |

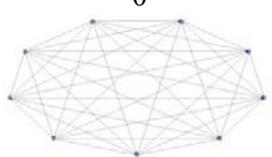

| | |
|---|---|
| # of nodes | 9 |
| # of edges | 36 |
| # of triangles | 84 |
| # of structural holes | 0 |

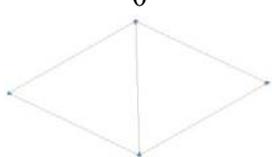 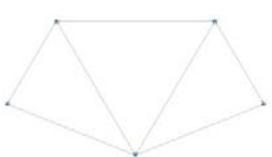 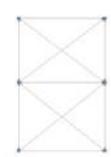

| | | | |
|---|---|---|---|
| # of nodes | 4 | 5 | 6 |
| # of edges | 5 | 7 | 11 |
| # of triangles | 2 | 3 | 8 |
| # of structural holes | 2 | 5 | 8 |

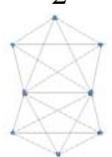 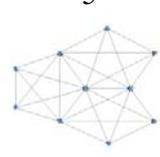 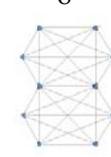

| | | | |
|---|---|---|---|
| # of nodes | 8 | 10 | 10 |
| # of edges | 19 | 27 | 29 |
| # of triangles | 20 | 31 | 40 |
| # of structural holes | 18 | 42 | 32 |

$$\begin{cases} \text{triangles:} & x+4y+10z = 5,597,411 \\ \text{edges:} & 3x+6y+10z = 8,474,226 \\ \text{nodes:} & 3x+4y+5z = 5,358,175 \end{cases}$$

However, we cannot get all positive solutions $x$, $y$, and $z$ for the equations above. And we try other kinds of combinations of $k$-cliques, and the situations are the same. The problem of Gleeson's algorithm is that it over-produces triangles with the same number of edges when working on the large-scale network.

6.5 A generalized version of Gleeson's algorithm

The only way out is that we should not constrain ourselves on $k$-clique. We can seek the combination of other motifs which have the following structures: i) there are three or more nodes in the motif; ii) edges are not completely connected between nodes in the motif as in a k-clique; and iii) therefore there are both triangles and structural holes in the motif.

For example, we can suppose the expected clustered random graph is composed of three motifs as showed in Figure 6: two triangles sharing a common edge, three triangles in a pentagon (a ring of 5 nodes) sharing a common node, and two 4-cliques sharing a common edge. And based on the numbers of nodes, edges, and triangles for each motif as showed in Table 1 we get

$$\begin{cases} \text{triangles:} & x+4y+8z = 5,597,411 \\ \text{edges:} & 3x+6y+11z = 8,474,226 \\ \text{nodes:} & 3x+4y+6z = 5,358,175 \end{cases}$$

And the solutions for the equations above is

$$\begin{cases} x = 395,455 \\ y = 270,343 \\ z = 515,073 \end{cases}$$

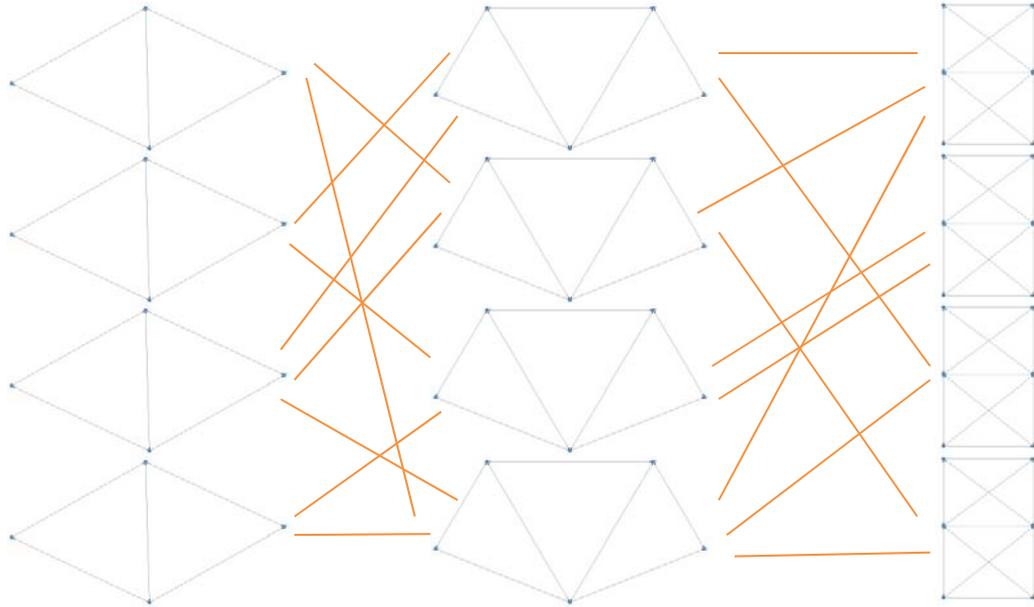

Figure 6. A clustered random graph formed by linking three motifs: two triangles sharing a common edge, three triangles in a pentagon sharing a common node, and two 4-cliques sharing a common edge

After the expected number of triangles is fit, we continue to add external links among the motifs. There are 15,913,611 edges is the graph and 15,233,033 of them are between the nodes whose nodal degree is greater than 2. There are 8,474,226 edges involving in triangles and thus the number of external links is 6,758,807 = 15,233,033 - 8,474,226.

And after randomly adding the 6,758,807 external links among motifs as in Figure 6 and also generating the edges incident to nodes of single degree, we get a clustered random graph with 6,719,330 nodes, 5,597,414 triangles, and 15,913,611 edges among which 8,474,226 are used for triangles and 6,758,807 are external links. However, we get only 81,181,872 2-paths which is much smaller than that in the real network, and the global clustering coefficient $C_\Delta$ is 0.21 and the average clustering coefficient $C(G)$ is 0.33.

The problem of the combination of three motifs as showed in Figure 6 – two triangles sharing a common edge, three triangles in a pentagon sharing a common node, and two 4-cliques sharing a common edge – is that we don't get enough 2-paths, which is identical to that we don't get enough structural holes since the number of triangles is fixed. There are 109,383,149 = 126,175,382 (the number of 2-paths) - 3×5,597,411 (the number of triangles) structural holes in the real network. And in the combination of three motifs mentioned above according to the statistics in Table 1 we already get

$2x + 5y + 8z = 2 \times 395,455 + 5 \times 270,343 + 8 \times 515,073 = 6,263,209$ structural holes inside the motifs, and we need 103,119,940 = 109,383,149 – 6,263,209 more structural holes among the motifs by adding 6,758,807 external links, which mean on average each external link will help to produce 15.26 = 103,119,940/6,758,807 structural holes between the nodes inside motifs. And external links are generated between nodes of motifs and thus we need the sum of nodal degrees of each pair of nodes to be 15.26 on average.

However, for the three motifs we use above, the maximum nodal degree is 5 which belong to the node involving in the sharing edge of two 4-cliques, and therefore the sum of nodal degrees of each pair of nodes is at most 10, which is much less than 15.26.

A combination of 7-clique, 8-clique, and 9-clique will help to make the range of sum of nodal degrees of each pair of nodes rise up to between 14 and 18 and 15.26 is just in this range. However, as we point out earlier, we cannot get all positive numbers of 7-clique, 8-clique, and 9-clique for generating the required number of triangles since the triangles are over-produced.

Now I am looking for a solution that can make the best of both worlds – to find some motifs consisting of both greater-nodal-degree nodes and greater number of structural holes. It turns out one of the methods is a combination of four motifs as showed in Figure 7 – *a*) two triangles sharing a common edge, *b*) three triangles in a pentagon sharing a common node, *c*) three 5-cliques sharing a common node, and *d*) two 6-cliques sharing a common edge, and both motif *c* and *d* have nodes with greater nodal degree and greater numbers of structural holes. And based on the statistics in Table 1 we get

$$\begin{cases} \text{triangles: } 2x + 3y + 31z + 40w = 5,597,411 \\ \text{edges: } \quad 5x + 7y + 27z + 29w = 8,474,226 \\ \text{nodes: } \quad 4x + 5y + 10z + 10w = 5,358,175 \end{cases}$$

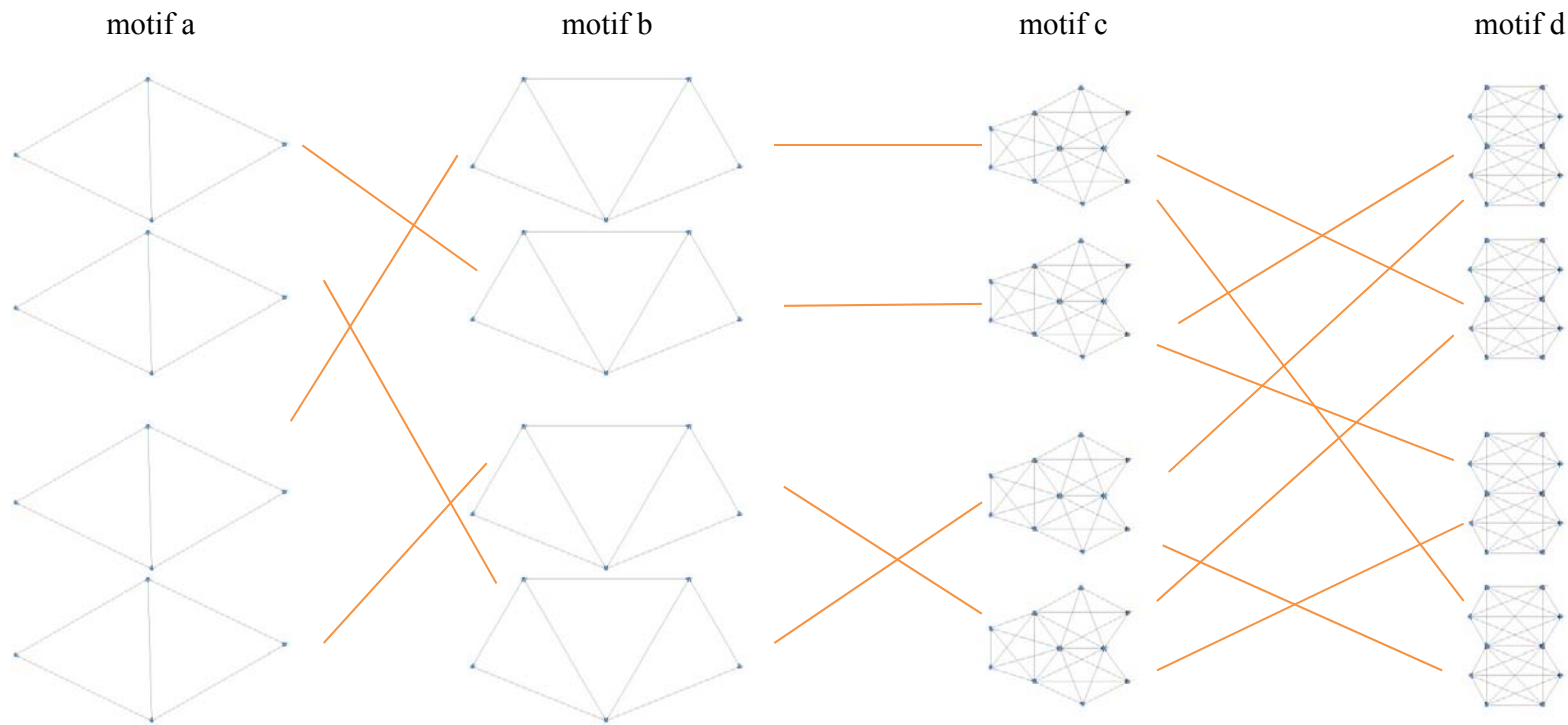

Figure 7. A clustered random graph formed by linking four motifs: a) two triangles sharing a common edge, b) three triangles in a pentagon sharing a common node, c) three 5-cliques sharing a common node, and d) two 6-cliques sharing a common edge

If we force that the number of motif *c* and that of motif *d* to be equal, we get

$$\begin{cases} x = 373,255 \\ y = 601,383 \\ z = 42,912 \\ w = 42,912 \end{cases}$$

There are 109,383,149 structural holes in the real network. And according to the table in the combination of four motifs above we already get

$2x + 5y + 42z + 32w = 2 \times 373,255 + 5 \times 601,383 + 42 \times 42,912 + 32 \times 42,912 = 6,928,913$ structural holes inside motifs, and we need 102,454,236 = 109,383,149 – 6,928,913 more structural holes between nodes of motifs by adding 6,758,807 external links, which mean on average each external link will help to generate 15.16 = 102,454,236/6,758,807 structural holes. And since the greater-nodal-degree nodes are located in motif *c* and *d*, we assign half external links between motif *c* and *d*, and one quarter each between motif *a* and *b* and between motif *b* and *c*.

In this way we get a clustered random graph of 6,719,330 nodes, 15,913,611 edges, 5,597,411 triangles, 109,379,166 structural holes, and 126,171,399 = 3 × 5,597,411 + 109,379,166 2-paths. The global clustering coefficient $C_\Delta$ or transitivity measure $T(G)$ is 3 ×5,597,411 / 126,171,399 = 0.13, which is the same as in the real network. The average clustering coefficient of the clustered random graph is 0.35, which is greater than that in the real network 0.24.

The algorithm takes about 9 hours to get the expected clustered random graph on the same server we use for the former three algorithms.

7. Randomness evaluation of the algorithms for generating clustered random graphs

I have mentioned that in a large-scale Erdős-Rényi random graph where the number of nodes $|V|$ at the nodal level and the number of edges $|E|$ at the dyadic level are kept, the tie formation probability $p$ is equal to the network density $\Delta$ as the network grows larger and larger. Therefore we can use the density value to evaluate the independence of an edge from the existence of any other ones.

For the clustered random graph we add one more constraint at the triadic level: the global clustering coefficient $C_\Delta$ or transitivity measure $T(G)$. We have got three kinds of clustered random graphs in which the global clustering coefficient $C_\Delta$ or transitivity measure $T(G)$ is the same as in the real network based on the algorithm of Guo and Kraines, the algorithm of Bansal et al., and the generalized version of Gleeson's algorithm. The network density $\Delta$ is 7.05e-7 in the real network. And as a result to evaluate the randomness of the clustered random graphs I need to look at the probabilities of edge existence of even higher-order configurations at the tetradic, pentadic, and/or hexadic levels and expect them to be close to the independent probability $p$ of tie formation which is equal to the network density $\Delta$ = 7.05e-7. Due to the limited RAM of the server (64GB) I only check the probabilities at the tetradic and pentadic levels and stop at the hexadic level.

As showed in Table 2, in the initiated random graph $G$ I generate for the rewiring processes by the algorithm of Guo and Kraines, at the triadic level the global clustering coefficient $C_\Delta$ or transitivity measure $T(G)$ is 7.50e-7 which is very close to the network density 7.05e-7, and at the tetradic and the pentadic levels the ratios of numbers of closed 4-/5- paths over numbers of both open and closed 4-/5- paths are both at the e-7 level, which confirms that at higher-order configuration levels the graph is random.

Table 2. Probability of edge existence at tetradic and pentadic levels

|  | Initiated random graph $G$ | Clustered random graph by algorithm of Guo and Kraines | Clustered random graph by algorithm of Bansal et al. | Clustered random graph by algorithm by the generalized version of Gleeson's algorithm |
|---|---|---|---|---|
| Triadic level ($C_\Delta / T(G)$) 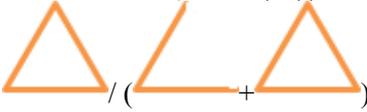 | 7.50e-7 | 0.13 | 0.13 | 0.13 |
| Tetradic level 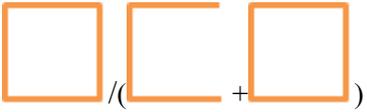 | 7.13e-7 | 5.15e-4 | 4.17e-4 | 2.66e-6 |
| Pentadic level 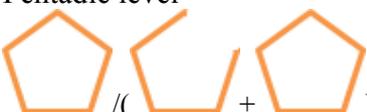 | 5.69e-7 | 3.87e-5 | 3.83e-5 | 6.44e-6 |

By adopting the algorithm of Guo and Kraines, the algorithms of Bansal et al, and the generalized version of Gleeson's algorithm, I generate three sets of clustered random graphs which successfully reproduce the global clustering coefficient $C_\Delta$ or transitivity measure $T(G)$ 0.13 as in the real network. However, as showed in Table 2, in the clustered random graph generated by the generalized version of Gleeson's algorithm the probabilities of edge existence at the tetradic and pentadic levels are at the $e^{-6}$ level which much closer to the network density than those generated by the other two algorithms. Therefore the graph generated by the generalized version of Gleeson's algorithm is more random than the other two.

8. Conclusions

Random graph is commonly used to compare with the real network. However, a Erdős-Rényi random graph in large-scale often lacks of the local structure beyond the dyadic level and as a result we need to generate the clustered random graph instead of the Erdős-Rényi random graph to compare the local structure at the triadic level.

Table 3. Algorithm summary for generating large-scale clustered random graphs

| | | Algorithm of Guo and Kraines | Algorithm of Bansal et al. | Newman's algorithm | The generalized version of Gleeson's algorithm |
|---|---|---|---|---|---|
| Nodal level | # of nodes | √ | √ | √ | √ |
| | nodal degree for each node[8] | × | √ | × | × |
| Dyadic level | # of edges | √ | √ | √ | √ |
| | Average nodal degree | √ | √ | √ | √ |
| | Network density | √ | √ | √ | √ |
| Triadic level | # of 2-paths | × | √ | × | √ |
| | # of structural holes | × | √ | × | √ |
| | # of triangles | × | √ | √ | √ |
| | Global clustering coefficient | √ | √ | × | √ |
| | Average clustering coefficient | × | × | × | × |

As showed in Table 3, we get three kinds of clustering random graphs in which the global clustering coefficient $C_\Delta$ or transitivity measure $T(G)$ as well as the number of nodes $|V|$ and the number of edges $|E|$ are the same as in the real network based on the algorithm of Guo and Kraines, the algorithm of Bansal et al., and the generalized version of Gleeson's algorithm. The

---

[8] To preserve the nodal degree of each node in the real network is necessary for the algorithm of Bansal et al. since only in this way the number of 2-paths can be fixed and we can just keep rewiring until we get the expected number of closed 2-paths – triangles. But it is not necessary for the Newman's algorithm and the generalized version of Gleeson's algorithm which use generating function to reproduce the expected numbers of triangles and structural holes.

Newman's algorithm doesn't work because it over-uses edges to generate the same number of triangles in the real network and as a result it does not persist the number of 2-paths/structural holes and the global clustering coefficient is not kept.

And by comparing the probability of edge existence at the tetradic and pentadic levels, the clustered random graph generated by the generalized version of Gleeson's algorithm seems to be more random than those generated by the algorithm of Guo and Kraines and the algorithm of Bansal et al.

Another advantage of the generalized version of Gleeson's algorithm is its computation time. While it takes weeks to generate the clustered random graph by the algorithm of Guo and Kraines and that of Bansal et al., we can get a clustered random graph based on the generalized version of Gleeson's algorithm in nine hours.

One criticism to Gleeson's algorithm is that it might not be a random process that nodes are set to be clustered in $k$-cliques to generate triangles. This critique could also be applied to the generalized version of Gleeson's algorithm – it might not be a random process that nodes are set to be clustered in motifs to generated triangles, and of course also applied to the specific version of Gleeson's algorithm – Newman's algorithm. However, from this perspective both the algorithm of Guo and Kraines and that of Bansal et al. do not have any advantage since the rewiring process might not be random either.

From my point of view, as long as we fix the global clustering coefficient as in the real network at the triadic level, the generation process is no longer as random as supposed in the critique. What we can assure is that each node has the same opportunity to be assigned to a configuration – a triangle as in Newman's algorithm, a $k$-clique as in Gleeson's algorithm, and a motif as in the generalized version of Gleeson's algorithm – or to a rewiring process as in the algorithm of Guo and Kraines and that of Bansal et al. And at the even higher tetradic and pentadic levels, the tie formation process is inclined to be random.

# References

Bagrow, J. P., Wang, D., & Barabási, A.-L. (2011). Collective Response on Human Populations to Large-scale Emergencies. *PLoS One*, **6**, 1-8.

Bansal, S., Khandelwal, S., & Meyers, L. A. (2009). Exploring biological network structure with clustered random networks. *BMC Bioinformatics*, **10**, 405-419.

Barabási, A.-L., & Albert, R. (1999). Emergence of Scaling in Random Networks. *Science*, **286**, 509-512.

Burt, R. S. (1995). *Structural Holes: The Social Structure of Competition*. Cambridge, MA: Harvard University Press.

Cartwright, D. & Harary, F. (1956). Structural Balance: A Generalization of Heider's Theory. *Psychological Review*, **63**, 277-293.

Davis, J. A. (1967). Clustering and Structural Balance in Graphs. *Human Relations*, **30**, 181-187.

Davis, J. A. (1979). The Davis /Holland /Leinhardt Studies: An Overview. In Holland, P. W., & Leinhardt, S. (Eds.), *Perspectives on Social Network Research* (pp. 51-62). New York: Academic Press.

Ercsey-Ravasz, M., Lichtenwalter, R. N., Chawla, N. V., & Toroczkai, Z. (2011). Range-limited Centrality Measures in Non-weighted and Weighted Complex Networks, *arXiv* e-print, **1111.5382**.

Erdős, P., & Rényi, A. (1959). On Random Graphs I. *Publicationes Mathematicae*, **6**, 290-297.

Ghoshal, G., & Barabási, A.-L. (2011). Ranking Stability and Super-stable Nodes in Complex Networks. *Nature Communications*, **2**, 1-7.

Gleeson, J. P. (2009). Bond Percolation on a Class of Clustered Random Networks. *Physical Review E.*, **80**, 036107.

Granovetter, M. (1973). The Strength of Weak Ties. *American Journal of Sociology*, **78**, 1360-1380.

Guo, W., & Kraines, S. B. (2009). A Random Network Generator with Finely Tunable Clustering Coefficient for Small-world Social Networks. In *Proceedings of the 2009 International Conference on Computational Aspects of Social Networks* (pp. 10-17). Washington, DC: IEEE Computer Society.

Heider, F. (1946). Attitudes and Cgnitive Organization. *Journal of Psychology*, **21**, 107-112.

Hidalgo, C. A., & Rodriguez-Sickert, C. (2008). The Dynamics of a Mobile Phone Network. *Physica A*, **387**, 3017-3024.

Krackhardt, D. (1998). Simmelian Ties: Super Strong and Sticky. In R. M. Kramer, & M. A. Neale (Eds.), *Power and Influence in Organizations* (pp. 21-38). Thousand Oaks, CA: Sage.

Krackhardt, D., & Handcock, M. S. (2006). Heider vs Simmel: Emergent Features in Dynamic Structures. In E. M. Airoldi, & D. M. Blei (Eds.), *Statistical Network Analysis: Models, Issues and New Directions (ICML 2006)* (pp.14-27). Berlin: Springer.

Lichtenwalter, R. N., Lussier, J. T., & Chawla, N. V. (2010). New Perspectives and Methods in Link Prediction. In *Proceedings of the 16th ACM SIGKDD International Conference on Knowledge Discovery and Data Mining (KDD)* (pp.243-252). New York: ACM.

Liu, Y.-Y., Slotine, J.-J., & Barabási, A.-L. (2011). Controllability of Complex Networks. *Nature*, **473**, 167-173.

Newman, M. E. J. (2009). Random Graphs with Clustering. *Physical Review Letters*, **103**, 05870.

Onnela, J. P., Arbesman, S., Gonzalez, M. C., Barabási, A.-L., & Christakis, N. A. (2011). Geographic Constraints on Social Network Groups. *PLoS One*, **6**, 1-7.

Opsahl, T., Colizza, V., Panzarasa, P., & Ramasco, J. J. (2008). Prominence and Control: The Weighted Rich-club Effect. *Physical Review Letters*, **101**, 168702.

Raeder, T., Lizardo, O., Hachen, D., & Chawla, N. V. (2011). Predictors of Short-term Decay of Cell Phone Contacts in a Large-scale Communication Network. *Social Networks*, **33**, 245-257.

Simmel, G. (1908/1950). *The Sociology of Georg Simmel*. New York: Free Press.

Travers, J., & Milgram, S. (1969). An Experimental Study of the Small World Problem. *Sociometry*, **32**, 425-443.

Wang, D., Pedreschi, D., Song, C, Giannotti, F., & Barabási, A.-L. (2011). Human Mobility, Social Ties, and Link Prediction. In *ACM SIGKDD International Conference on Knowledge Discovery and Data Mining (KDD)* (pp. 1100-1108). New York: ACM.

Wasserman S., & Faust K. (1994). *Social Network Analysis: Methods and Applications*. New York: Cambridge University Press.

Watts, D. J., & Strogatz, S. H. (1998). Collective Dynamics of "Small-world" Networks. *Nature*, **393**, 440-442.